\documentclass[twocolumn,showpacs,amsfonts,aps,prc,nofootinbib,floatfix,superscriptaddress]{revtex4}

\usepackage{bm}
\usepackage{graphicx}
\usepackage{amsmath}
\usepackage{epstopdf}
\usepackage{xcolor}
\usepackage[normalem]{ulem}

\def\l{\left}
\def\r{\right}
\def\la{\langle}
\def\ra{\rangle}
\def\dla{\la\!\la}
\def\dra{\ra\!\ra}

\def\ebsT{\l( \eta/s \r)(T)}

\def\eq{{\,=\,}}
\def\ev{{\mathrm{ev}}}

\newcommand{\avg}[1]{\bigl\langle #1 \bigr\rangle_S} 
\newcommand{\mean}[1]{\bigl\langle #1\bigr\rangle} 

\definecolor{olive}{rgb}{0.3, 0.4, .1}
\definecolor{fore}{RGB}{249,242,215}
\definecolor{back}{RGB}{51,51,51}
\definecolor{title}{RGB}{255,0,90}
\definecolor{dgreen}{rgb}{0.,0.6,0.}
\definecolor{gold}{rgb}{1.,0.84,0.}
\definecolor{JungleGreen}{cmyk}{0.99,0,0.52,0}
\definecolor{BlueGreen}{cmyk}{0.85,0,0.33,0}
\definecolor{RawSienna}{cmyk}{0,0.72,1,0.45}
\definecolor{Magenta}{cmyk}{0,1,0,0}

\begin{document}

\title{Interferometric signatures of the temperature dependence of the specific shear viscosity in heavy-ion collisions}

\author{Christopher Plumberg}
\author{Ulrich Heinz}
\affiliation{Department of Physics, The Ohio State University,
  Columbus, OH 43210-1117, USA}

\begin{abstract}
Recent work has shown that a temperature dependence of the shear viscosity to entropy ratio, $\eta/s$, influences the collective flow pattern in heavy-ion collisions in characteristic ways that can be measured by studying hadron transverse momentum spectra and their anisotropies.  Here we point out that it also affects the pair momentum dependence of the Hanbury-Brown--Twiss (HBT) radii (the source size parameters extracted from two-particle intensity interferometry) and the variance of their event-by-event fluctuations.  This observation establishes interferometric signatures as useful observables to complement the constraining power of single-particle spectra on the temperature dependence of $\eta/s$.
\end{abstract}


\date{\today}

\maketitle

\section{Introduction}
\label{sec1}
\vspace*{-2mm}

Recently, a great deal of theoretical attention has been devoted to the accurate determination of the shear-viscosity-to-entropy-density ratio ($\eta/s$) of the quark-gluon plasma (QGP) and the hadron resonance gas (HRG), and how its temperature dependence around the transition point between these two phases might be probed in heavy-ion collisions.  In particular, a recent study \cite{Molnar:2014zha} suggested the possibility of discriminating experimentally between different parametrizations of the temperature dependence of $\eta/s$ using systematic analyses of the anisotropic flows $v_2$ and $v_4$ as functions of transverse momentum $p_T$ and pseudorapidity $\eta$.  In this paper, we present a complementary analysis of the HBT radii for Au+Au collisions at RHIC energies, to argue that the HBT radii can be used profitably to help constrain viable parametrizations of $\l(\eta/s\r)(T)$.

The HBT radii are physical observables which are well defined on an event-by-event basis.  As parameters that describe the size and shape of the particle emitting source, they fluctuate from event to event.  Experimental data so far \cite{Bearden:1998aq,Adamova:2002wi,Abelev:2013pqa,Adamczyk:2014mxp,Adare:2014qvs,Adam:2015pya} determine only the average source radii characterizing a large ensemble of events  \cite{Plumberg:2013nga}.  Elsewhere \cite{Plumberg:2015cjp} we discuss new ideas of how to complement such measurements of the ensemble-averaged HBT radii by a determination of their variances, i.e., the width parameters characterizing their event-by-event fluctuations.  In that work \cite{Plumberg:2015cjp} it is also shown that the experimentally accessible observables characterizing the mean source radii and their variances clearly track the algebraic mean and variance of the source radii that are calculated theoretically from the individual emission functions of a large set of dynamically simulated heavy-ion collisions with fluctuating initial conditions.  We here show the transverse pair momentum ($\vec{K}_T$) dependencies of these mean radii and their variances for the same four parametrizations of the specific shear viscosity $\eta/s$ studied in \cite{Molnar:2014zha}.  We also compare the mean radii to those one would obtain from a single average source by dynamically evolving a single initial profile, constructed as the ensemble average of the fluctuating initial conditions of the full ensemble.

To save time and effort, we employ for this pioneering study the shortcut of calculating the HBT radii from the source variances of the emission function \cite{Heinz:1999rw,Kolb:2003dz,Wiedemann:1999qn,Lisa:2008gf} for directly emitted pions only, instead of performing a 3-D Gaussian fit to the correlation function including all resonance decays \cite{Lisa:2008gf,Kisiel:2006is,Romatschke:2007jx,Bozek:2009ty}.  This approximation, which holds exactly only for Gaussian sources \cite{Heinz:1999rw}, is known to be sufficient to discern qualitative features of the HBT radii and their $K_T$ dependences, although it is not accurate enough for quantitative comparisons with experimental data \cite{Frodermann:2006sp}.

\section{Analysis}
\label{sec2}
\subsection{Initial conditions}
\label{ssec2a}

In this paper, we compute the HBT radii, as outlined in the previous section, for central (0-10\%) Au+Au collisions at 200\,$A$\,GeV, based on charged pion correlations.  We use the iEBE-VISHNU package \cite{Shen:2014vra} in modeling the evolution of heavy-ion collisions.  For the (hydrodynamical) initial conditions of the fireball, we employ the MC-Glauber model \cite{Alver:2008aq} with p+p multiplicity fluctuations \cite{Shen:2014vra} to generate the initial entropy density distribution in the transverse plane.  The subsequent dynamical evolution is performed on an event-by-event basis, meaning that each set of initial conditions that we generate initializes one of the $N_{\ev}$ hydrodynamically evolved events in our ensemble.  In this paper, we take $N_{\ev}\eq 1000$.

\subsection{Hydrodynamic evolution and $(\eta/s)(T)$}
\label{ssec2b}

We perform the hydrodynamical evolution of the initial conditions with the VISH2+1 code \cite{Song:2007ux}, using the s95p-PCE165-v0 equation of state \cite{Shen:2010uy}.  We choose this evolution to begin at an initial proper time $\tau_0\eq0.6$\,fm/$c$, where $\tau\eq0$ corresponds to the instant at which the two nuclei collide.  Also, although necessary for quantitatively precise comparisons with experimental data \cite{Pratt:2008qv}, we do not incorporate here any initial transverse flow into the hydrodynamical evolution.  

\begin{figure}[hb!]
\includegraphics[scale=0.65]{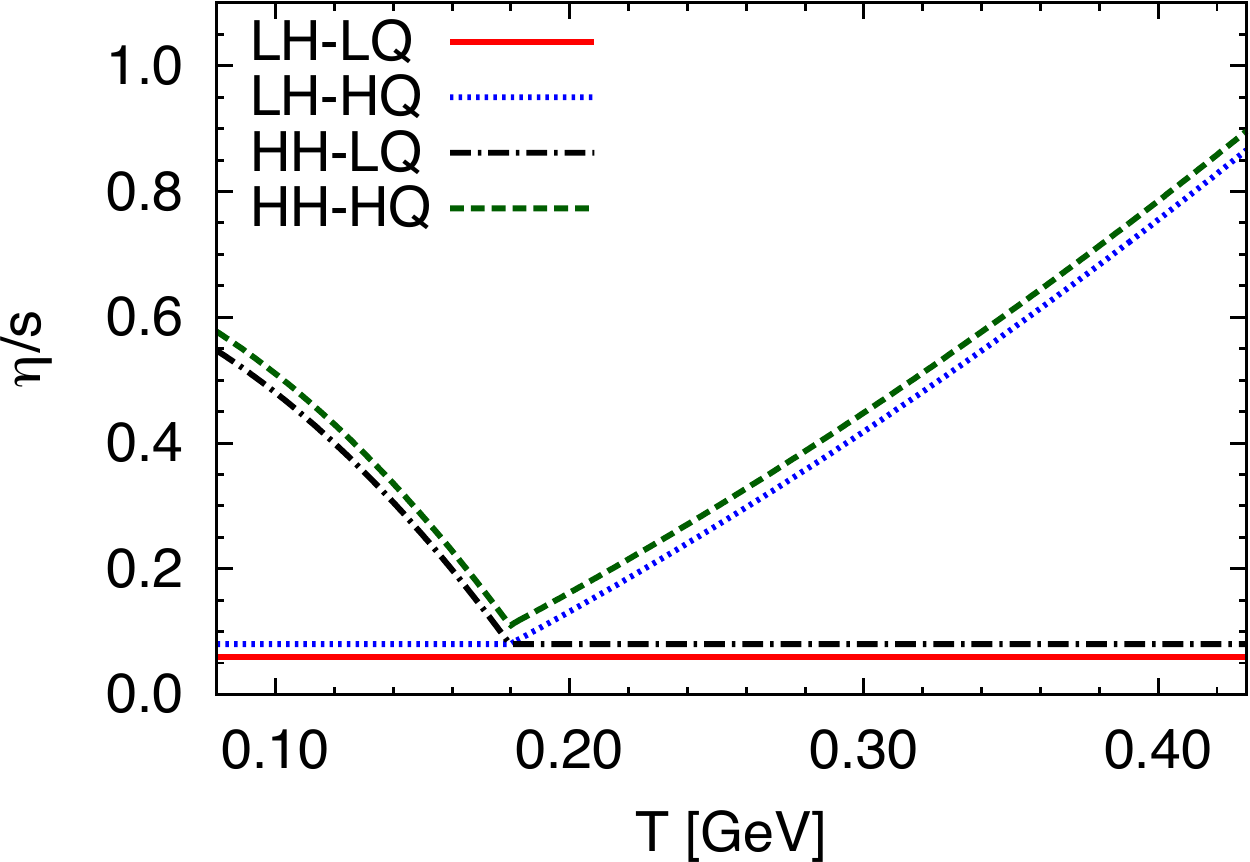}
\caption{The different parametrizations of $(\eta/s)(T)$ used in this work, taken from \cite{Molnar:2014zha}.  
\label{F1}
}
\end{figure}

For our study, we use for $\l(\eta/s\r)(T)$ one of the four parametrizations given in \cite{Molnar:2014zha}:
\begin{itemize}
\item LH-LQ, in which $(\eta/s)(T)=0.08$ for all $T$;

\item LH-HQ, in which $(\eta/s)(T)=0.08$ for $T \leq T_{\mathrm{tr}}$ and
\begin{eqnarray}
(\eta/s)(T)_{\mathrm{QGP}} &=& -0.289+0.288\,\frac{T}{T_{\mathrm{tr}}}
          +0.0818\left( \frac{T}{T_{\mathrm{tr}}}\right)^2,\nonumber\\
         & \quad & {\mbox{for $T > T_{\mathrm{tr}}$;}}
\end{eqnarray}

\item HH-LQ, in which $(\eta/s)(T)=0.08$ for $T > T_{\mathrm{tr}}$ and
\begin{eqnarray}
(\eta/s)(T)_{\mathrm{HRG}} &=& 0.681-0.0594\,\frac{T}{T_{\mathrm{tr}}}
          -0.544\left( \frac{T}{T_{\mathrm{tr}}}\right)^2 \nonumber\\
          & \quad & {\mbox{for $T \leq T_{\mathrm{tr}}$;}}
\end{eqnarray}

\item HH-HQ, for which we use $(\eta/s)(T)_{\mathrm{HRG}}$ for $T \leq T_{\mathrm{tr}}$ and
  $(\eta/s)(T)_{\mathrm{QGP}}$ for $T > T_{\mathrm{tr}}$.
\end{itemize}  
Here, $T_{\mathrm{tr}} = 180$\,MeV \cite{Molnar:2014zha} represents the transition temperature between the QGP and HRG phases. The four parametrizations are shown in Fig. \ref{F1}.

As also pointed out in \cite{Molnar:2014zha}, for identical initial conditions viscous entropy production generates different final multiplicities for different choices of $\ebsT$. This effect is especially severe for the parametrizations LH-HQ and HH-HQ where at early times, when the longitudinal-transverse velocity shear is biggest, $\ebsT$ is very large. To compensate for this effect, we rescale our initial entropy profiles by a constant factor $\kappa_\mathrm{fit}$, which depends on the chosen parametrization of $\ebsT$.  For a given choice of $\ebsT$, $\kappa_\mathrm{fit}$ is chosen so that an ensemble-averaged initial entropy profile always reproduces the same final thermal $\pi^+$ multiplicity.\footnote{%
    Since we always decouble at the same constant freeze-out temperature, $\frac{dN_{\pi^+}}{d\eta}$ 
    is a fixed fraction of the total multiplicity $\frac{dN_\mathrm{ch}}{d\eta}$. Therefore, this normalization 
    ensures that, on average, all events have the same final charged multiplicity.}
Normalizing all of the averaged profiles to the ideal fluid case (which thus corresponds to $\kappa_\mathrm{fit}\eq1$) we find
\begin{itemize}
	\item LH-LQ: $\kappa_\mathrm{fit} = 0.889$
	\item LH-HQ: $\kappa_\mathrm{fit} = 0.596$
	\item HH-LQ: $\kappa_\mathrm{fit} = 0.878$
	\item HH-HQ: $\kappa_\mathrm{fit} = 0.589$
\end{itemize}%
As expected, parametrizations with a large plasma viscosity require a significant suppression of the initial entropy density profile in order to compensate for viscous heating and to reproduce the measured final multiplicities in Au+Au collisions at RHIC energies. Finally, we terminate the hydrodynamical evolution on a decoupling surface (the "freeze-out surface") of constant temperature $T_\mathrm{dec} = 120$ MeV.

\subsection{Pion emission function at freeze-out}
\label{ssec2c}

On the freeze-out surface for each event, the pion emission function is defined by the Cooper-Frye integral \cite{Cooper:1974mv,Schlei:1992jj,Chapman:1994xa}
\begin{eqnarray}
S(x,p) &=& \frac{1}{(2\pi)^3} \int_{\Sigma} p \cdot d^3 \sigma(y)\, \delta^4 (x{-}y)\, f(y,p)\,, \label{cooper_frye_defn1}
\\
f(x,p) 	&=& f_0 \l(x,p\r) + \delta f \l(x,p\r)\nonumber\\
		&=& \frac{1}{e^{(p \cdot u{-} \mu)/T}{-}1} + \frac{p^{\mu} p^{\nu} \pi_{\mu\nu} }{2 T^2 (e{+}{\cal P})} f_0 (1{+}f_0). 
\label{cooper_frye_defn2}
\end{eqnarray}  
Here, $\delta f$ is the first-order viscous corrections to the ideal distribution function $f_0$ \cite{Teaney:2003kp,Dusling:2009df} for which we assume a quadratic dependence on $p$. $\pi_{\mu\nu}(x)$ and $u^{\mu}(x)$ are the shear stress and flow velocity profiles along the freeze-out surface $\Sigma$, respectively. $\mu$, $T$, $e$, and $\cal P$ are the chemical potential, decoupling temperature, energy density, and pressure, respectively, which are all constant on $\Sigma$ by construction. $d^3\sigma_{\mu}(x)$ is the outward pointing normal vector on $\Sigma$ at point $x$.

\subsection{Ensemble averaging}
\label{ssec2d}

Theoretically, the HBT radii may be defined for either a single event or an ensemble containing many events. Due to statistical limitations arising from the finite number of particles emitted in a single event, precise measurements of the full set of HBT radii and their $K_T$ dependences for a single event are not possible \cite{Plumberg:2013nga}.  Three-dimensional experimental HBT analyses, therefore, exclusively report measurements based on large collections of events, rather than measurements of event-by-event HBT radii.  In order to obtain an apples-to-apples comparison with experimental data, theoretical HBT calculations must therefore, at some level, average over all events in the ensemble.

This procedure, known as \textit{ensemble averaging}, can be performed in several ways, two of which we consider here.  According to one prescription, the initial conditions for all events in the ensemble are averaged \textit{before} their hydrodynamical evolution.  We refer to this method of ensemble averaging as "single-shot hydrodynamics" (SSH) \cite{Qiu:2011iv}.  For a sufficiently large ensemble, averaging over many, individually fluctuating and bumpy initial conditions results in a \textit{smooth} initial transverse entropy density profile, which in turn may be evolved hydrodynamically as a single, averaged event.  This prescription eliminates any sensitivity to event-by-event fluctuations in the initial state.  We will here denote the HBT radii derived from single-shot hydrodynamics by $\bar{R}^2_{ij}\,(i,j = o,s,l)$.

An alternative prescription for computing the ensemble-averaged HBT radii evolves each fluctuating event \textit{independently} and averages over the entire ensemble only \textit{after} the HBT radii have been computed for each event from its individual hydrodynamic emission function $S(x,K)$.  This leads to the following definition of ensemble-averaged HBT radii:
\begin{equation}
\mean{R^2_{ij}} \equiv \frac{1}{N_{\ev}} \sum^{N_{\ev}}_{k=1} \l( R^2_{ij} \r)_{k}. \label{DEAdefn}
\end{equation} 
We denote by $\l( R^2_{ij} \r)_{k}$ the HBT radii of the $k$th event, and we label the mean HBT radii computed according to this "direct ensemble average" (DEA) by $\mean{R^2_{ij}}$.

\subsection{HBT calculations}
\label{ssec2e}

The basic formalism describing our calculation of the HBT radii was presented in \cite{Plumberg:2013nga}.  Once the emission function \eqref{cooper_frye_defn1} for a particular event has been obtained from the hydrodynamic output using Eq.~\eqref{cooper_frye_defn2}, the HBT radii corresponding to that emission function may be defined by \cite{Heinz:1999rw, Wiedemann:1999qn}
\begin{equation}
R^2_{ij}(\vec{K}) = \avg{(\tilde{x}_i - \beta_i \tilde{t})(\tilde{x}_j - \beta_j \tilde{t})},  \label{svHBT_defn}
\end{equation} 
where $\tilde{x}_{\mu}\eq{x}_{\mu}{-}\avg{x_{\mu}}$, $\vec{\beta}\eq\vec{K}/E_K$, $E_K\eq\sqrt{m_{\pi}^2{+}\vec{K}^2}$, and the $\vec{K}$ dependence of $R^2_{ij}$ arises from the $\vec{K}$ dependence of the emission function in the definition of the source average:
\begin{equation}
\avg{f(x)} \equiv \frac{\int d^4 x\, f(x)\, S(x,K)}{\int d^4 x\, S(x,K)}.  
\label{source_integral}
\end{equation}
$\vec{K} = (K_T, \Phi_K, Y)$ is the average pair momentum in the lab frame, $\vec{K}\eq\frac{1}{2} \l(\vec{p}_1{+}\vec{p}_2 \r)$.  In this paper, we consider only pairs of identical particles at mid-rapidity ($Y\eq0$). 

The dependence of the HBT radii on the pair emission angle $\Phi_K$ also allows them to be expanded in a Fourier series (we suppress the additional dependence of all HBT parameters on $K_T$ and $Y$),
\begin{eqnarray}
R^2_{ij}(\Phi_K) &=& R^2_{ij,0} + 2 \sum^{\infty}_{n=1} \l[ R^{2(c)}_{ij,n} \cos\bigl(n(\Phi_K{-}\Psi_n)\bigr) \right. 
\nonumber\\
	& & \qquad\qquad \left. +\ R^{2(s)}_{ij,n} \sin\bigl(n(\Phi_K{-}\Psi_n)\bigr) \r],
\label{R2ij_FT_defn}
\end{eqnarray}
where the flow-plane angle $\Psi_n$ is defined by the complex phase of the $n$th-order anisotropic flow coefficient $v_n$ \cite{Heinz:2013th}:
\begin{eqnarray}
\label{vn_defn}
  v_n e^{i n \Psi_n} &\equiv& \dla e^{in\Phi}\dra
\\ \nonumber
  &=& \frac{\int^{\infty}_0 dK_T K_T\int_{-\pi}^{\pi}d\Phi_K\, e^{i n \Phi_K}\int d^4 x \, S(x,K)}{\int^{\infty}_0 dK_T K_T\int_{-\pi}^{\pi}d\Phi_K\,\int d^4 x \, S(x,K)}. 
\end{eqnarray}
In this work, we average $R^2_s$, $R^2_o$, and $R^2_l$ over $\Phi_K$, but we report the 2nd $\Phi_K$-harmonic $R^2_{os,2}$ of the out-side cross-term $R^2_{os}(\Phi_K)$, since the mean value of its $\Phi_K$-average $\mean{R^2_{os,0}}$ vanishes for symmetric nucleus-nucleus collisions at midrapidity \cite{Heinz:2002au}.  For single events with fluctuating initial conditions, $R^2_{os,2}$ receives non-zero contributions from both $R^{2(c)}_{os,2}$ and $R^{2(s)}_{os,2}$ on an event-by-event basis, but the former contribution averages to zero for a sufficiently large ensemble. We therefore consider only the mean and variance of the event-wise fluctuations of $R^{2(s)}_{os,2} \equiv R^{2}_{os,2}$ for the remainder of this paper.  For notational simplicity, we will drop the '0' subscript from the azimuthally averaged radii, and write simply $R^2_{ij}\ (i,j\eq{o},s,l)$.

At $Y\eq0$, Eq.~(\ref{svHBT_defn}) gives
\begin{equation}
  R^2_s = \avg{\tilde{x}_s^2},\quad
  R^2_l = \avg{\tilde{x}_l^2}. 
  \label{R2sl_defn}
\end{equation}
$R^2_s$ and $R^2_l$ thus depend only on the geometric aspects of the emission function. By contrast, both $R^2_o$ and $R^2_{ol}$ involve a combination  of spatial and temporal source variances:
\begin{eqnarray}
R^2_o &=& \avg{\tilde{x}_o^2} - 2 \beta_T \avg{\tilde{x}_o \tilde{t}} + \beta_T^2 \avg{\tilde{t}^2} \label{R2o_defn}, \\
R^2_{os} &=& \avg{\tilde{x}_o \tilde{x}_s} - \beta_T \avg{\tilde{x}_s \tilde{t}}
\label{R2os_defn}
\end{eqnarray}
In \cite{Plumberg:2015cjp} we describe a method for experimentally measuring, in addition to the mean, also the variance of an event-by-event distribution of the HBT radii. As a prediction for such measurements, we here also compute the variances of the HBT radii for our ensemble of $N_{\ev} = 1000$ events, defined by%
\footnote{%
      The factor $\frac{1}{N_{\ev}{-}1}$ makes $\sigma^2_{ij}$ an unbiased estimator of the 
      true variance of the event-by-event distribution of HBT radii \cite{Tamhane:2000sda}.}
\begin{equation}
   \sigma^2_{ij} = \frac{1}{N_{\ev}{-}1} \sum^{N_{\ev}}_{k=1} 
   \l( \l( R^2_{ij} \r)_{k} - \mean{R^2_{ij}} \r)^2.
\end{equation} 
In this paper, we will use and plot $\sigma_{ij} / \mean{R^2_{ij}}$ as a measure for the widths of the HBT radii distributions.  For the quantity $R^{2}_{os,2}$, we define the corresponding variance to be
\begin{equation}
   \sigma^2_{os,2} = \frac{1}{N_{\ev}{-}1} \sum^{N_{\ev}}_{k=1} 
   \l( \l( R^2_{os,2} \r)_{k} - \mean{R^2_{os,2}} \r)^2.
\end{equation} 

\section{Results}
\label{sec3}
\subsection{Mean HBT radii and their $K_T$ dependence}
\label{sec3a}

In this subsection, we present the mean HBT radii computed according to the two averaging prescriptions described above (single-shot hydrodynamics (SSH) and the direct ensemble average (DEA)), and study their sensitivity to the temperature dependence of the specific shear viscosity $\ebsT$.

\begin{figure}[t!]
\hspace*{-3.5mm}\includegraphics[width=1.05\linewidth]{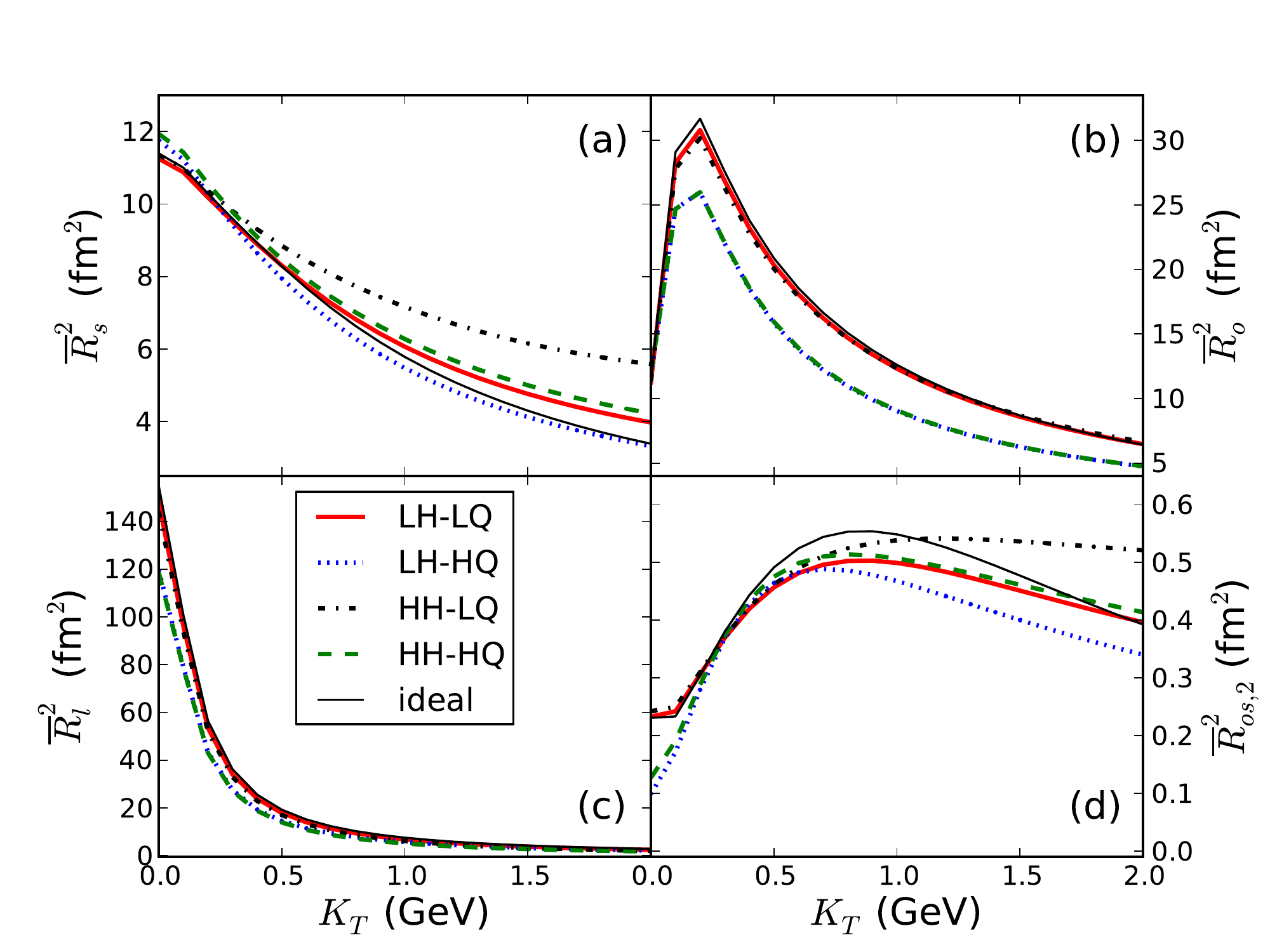}
\caption{ $\bar{R}^2_{ij}$ vs. $K_T$ from single-shot hydrodynamics, for ideal and viscous fluids with the temperature dependent shear viscosities $\ebsT$ shown in Fig. \ref{F1}. Note the almost perfect overlap of the green (dashed) and blue (dotted) curves in panels (b,c).
\label{F2}}
\end{figure}

\begin{figure}[t!]
\hspace*{-3.5mm}\includegraphics[width=1.05\linewidth]{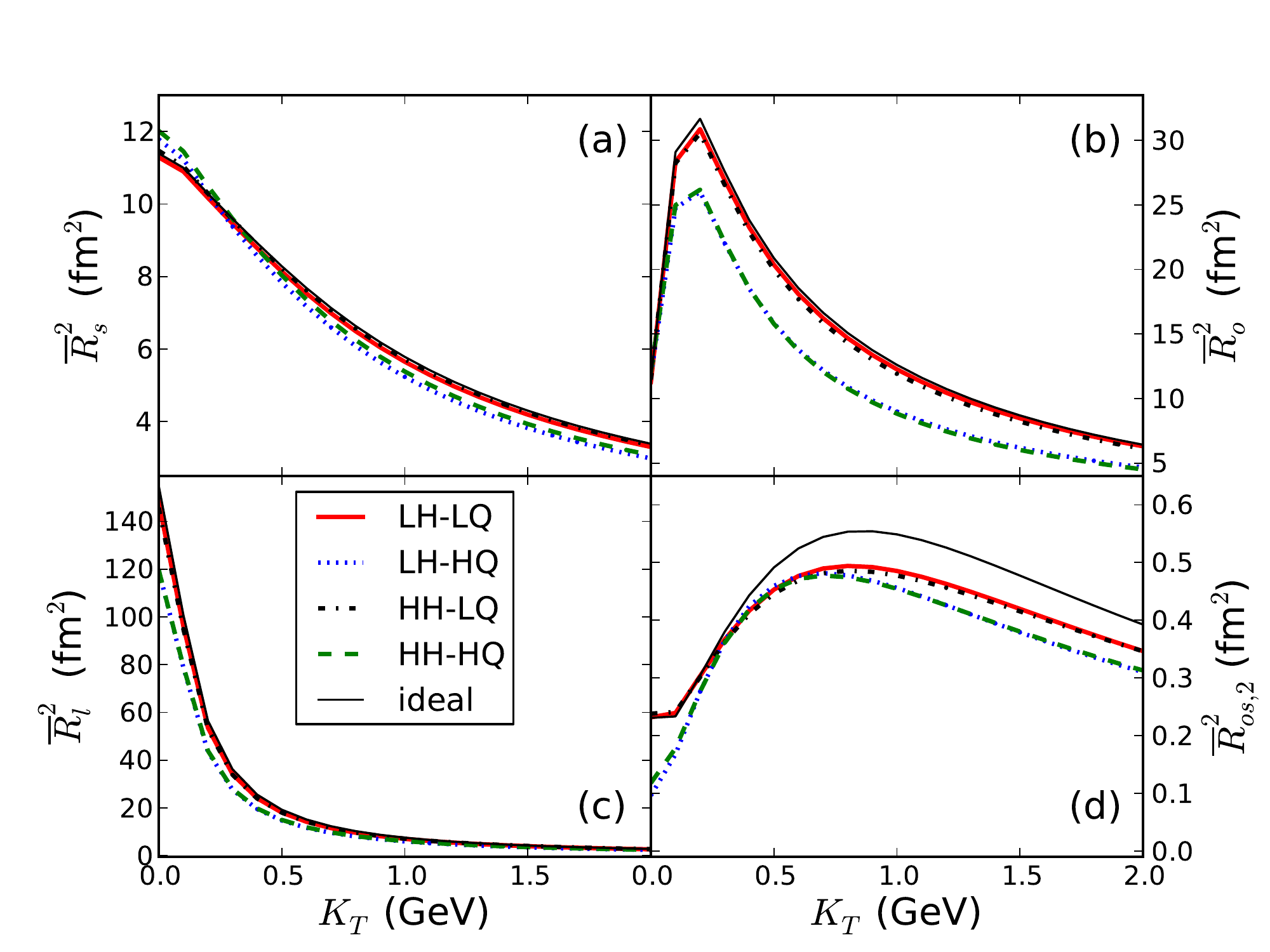}
\caption{Same as Fig.~\ref{F2}, but leaving out in Eq.~(\ref{cooper_frye_defn2}) the viscous correction $\delta f$ at freeze-out. Comparison of Figs.~\ref{F2}a,d and \ref{F3}a,d shows that the sensitivity of $\bar{R}^2_s(K_T)$ and $\bar{R}^2_{os,2}(K_T)$to the temperature dependence of $\ebsT$ below $T_\mathrm{tr}$ is almost entirely due to the theoretically poorly controlled viscous correction $\delta f$ at freeze-out, and that any sensitivity to the temperature dependence of $\ebsT$ during the dynamical evolution, i.e. {\em before freeze-out}, is strongly weighted at high temperatures, i.e. early times. The sensitivity to $\delta f$ at freeze-out of the HBT radii probing the outward and longitudinal dimensions of the emission function (panels (b,c)) is negligible. 
\label{F3}
}
\end{figure}

First we consider in Fig.~\ref{F2} the radii $\bar{R}^2_{ij}$ extracted from single-shot hydrodynamics which averages over event-by-event fluctuations in the initial state. Panels (a) and (d) show that the sideward radius $\bar{R}^2_s$ and the out-side cross term $\bar{R}^2_{os,2}$ exhibit sensitivity to the temperature dependence of $\ebsT$ both below and above $T_\mathrm{tr}$. However, $\ebsT$ affects hadronic observables in two distinct ways. First, there is the cumulative dynamical effect of shear viscosity on the development of flow in the fireball; flow probes the entire temperature history of $\ebsT$ between the initial and decoupling temperatures. Second, the emission function is affected by the viscous correction $\delta f$ to the distribution function at freeze-out, Eq.~(\ref{cooper_frye_defn2}), which is controlled by the value of $\ebsT$ on the freeze-out surface. This gives a contribution to observables such as the HBT radii that depends exclusively on the behavior of $\ebsT$ at $T_\mathrm{dec}$ and is basically independent of its prior history (except for the accumulated effect of $\ebsT$ on the flow pattern whose associated shear tensor also affects the shear stress at freeze-out). To separate the two effects, we plot in Fig.~\ref{F3} the HBT radii computed without the $\delta f$ correction, so that only the cumulative dynamical effects of $\ebsT$ remain. Figures~\ref{F3}a,d show that deleting $\delta f$ removes all sensitivity of the sideward and out-side HBT radii on the shear viscosity in the hadronic phase, leaving only a weak sensitivity on the behavior of $\ebsT$ in the high-temperature QGP phase.

On the other hand, comparison of panels (b,c) in Figs.~\ref{F2} and \ref{F3} reveals that the outward and longitudinal HBT radii are almost completely unaffected by the behavior of $\ebsT$ below $T_\mathrm{tr}$ and, in particular, receive no significant contribution from the viscous correction $\delta f$ at freeze-out. This finding is at variance with the results reported in Teaney's pioneering analysis \cite{Teaney:2003kp} of $\delta f$-effects on the HBT radii. We note, however, that Teaney's analysis was based on a blast-wave parametrization of the hydrodynamic flow profile at freeze-out, rather than the profile from a genuine hydrodynamic simulation of the flow. It was noted before in Appendix E of Ref.~\cite{Song:2007ux} that $\delta f$ is very sensitive to the details of the velocity shear tensor at freeze-out, and even its sign is fragile.
%
\begin{figure}[t!]
\hspace*{-3.5mm}\includegraphics[width=1.05\linewidth]{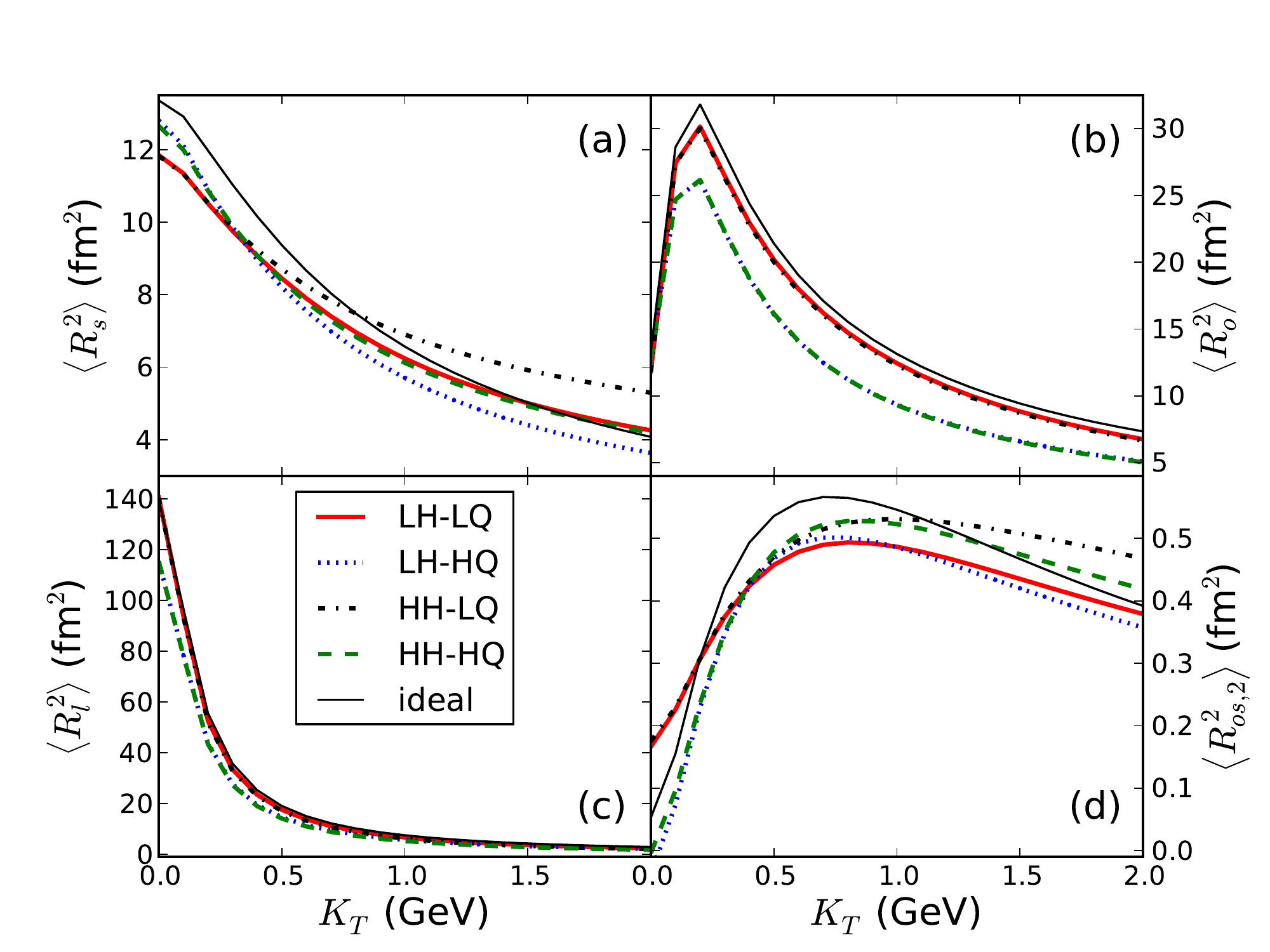}
\caption{ The ensemble averaged $\mean{R^2_{ij}}$ as a function of pair momentum $K_T$, for the same choices of $\ebsT$ as in Figs.~\ref{F2} and \ref{F3}. This figure should be compared with Fig.~\ref{F2}.
\label{F4}
}
\end{figure}
%
Furthermore, there are well-known conceptual uncertainties about the correct form of $\delta f$, especially its $p_T$ dependence \cite{Dusling:2009df}. These uncertainties render suspect any observable that strongly depends on $\delta f$ at freeze-out.%
\footnote{%
    We comment that $\delta f$-related uncertainties may also affect some of the 
    conclusions in Ref.~\cite{Molnar:2014zha} about the dependence of charged hadron
    elliptic and quadrangular flow, $v_2^\mathrm{ch}(p_T)$ and $v_4^\mathrm{ch}(p_T)$,
    on the behavior of $\ebsT$ in the hadronic phase. For example, we found that, when 
    calculated without the $\delta f$ correction, almost all sensitivity of $v_2(p_T)$ of directly 
    emitted (``thermal'') pions to the temperature dependence of $\ebsT$ in the hadronic 
    phase disappears, leaving only a (significantly weaker, but robust) dependence of this
    observable to the temperature dependence of $\ebsT$ in the QGP phase above 
    $T_\mathrm{tr}$.}
Therefore, we caution the reader not to trust the apparent sensitivity in Fig.~\ref{F2}a,d of $\bar{R}^2_s(K_T)$ and $\bar{R}^2_{os,2}(K_T)$ to the behavior of $\ebsT$ in the hadronic phase. On the other hand, the sensitivity of $\bar{R}^2_o(K_T)$ and $\bar{R}^2_l(K_T)$ exclusively to the temperature dependence of $\ebsT$ in the QGP phase {\em above} $T_\mathrm{tr}$ appears to be robust and unaffected by freeze-out uncertainties related to $\delta f$. These high-temperature effects of $\ebsT$ on $\bar{R}^2_{o}$ and $\bar{R}^2_{l}$ are not huge, but they can reach 20\% (for the squared HBT radii) for a range of $K_T$ values below 1.5\,GeV.

Studies such as \cite{Romatschke:2007jx,Bozek:2009ty} have explored the sensitivity of the HBT radii to shear viscosity using constant ($T$-independent) $\eta/s$ and varying that constant.  Different from our approach here, when changing $\eta/s$ they also changed other hydrodynamic parameters to ensure that not only the normalization but also the slope of the $p_T$-spectra was held fixed.  Additionally, both of these studies extract the HBT radii from a Gaussian fit to the full three-dimensional correlation function instead of using the short-cut through the source variances.  These differences make a direct comparison of our work with their results difficult.  Our findings are qualitatively supported by the work \cite{Gombeaud:2009fk} (whose authors, like us, used the source variances method) where it was found that an increase in the Knudsen number (or, equivalently, $\eta/s$) resulted in a decrease in $R^2_o$.  The purpose of the present work is not a realistic comparison with experimental data which, as done in \cite{Romatschke:2007jx,Bozek:2009ty}, would require a simultaneous tuning of several additional parameters (see also Ref.~\cite{Bernhard:2015hxa}).  Our goal is to check systematically the sensitivity of the HBT radii to a possible $T$-dependence of $\eta/s$, in particular in the context of event-by-event source fluctuations.  It is to these fluctuations that we turn our attention next.

\begin{figure}[t!]
\hspace*{-3.5mm}\includegraphics[width=1.05\linewidth]{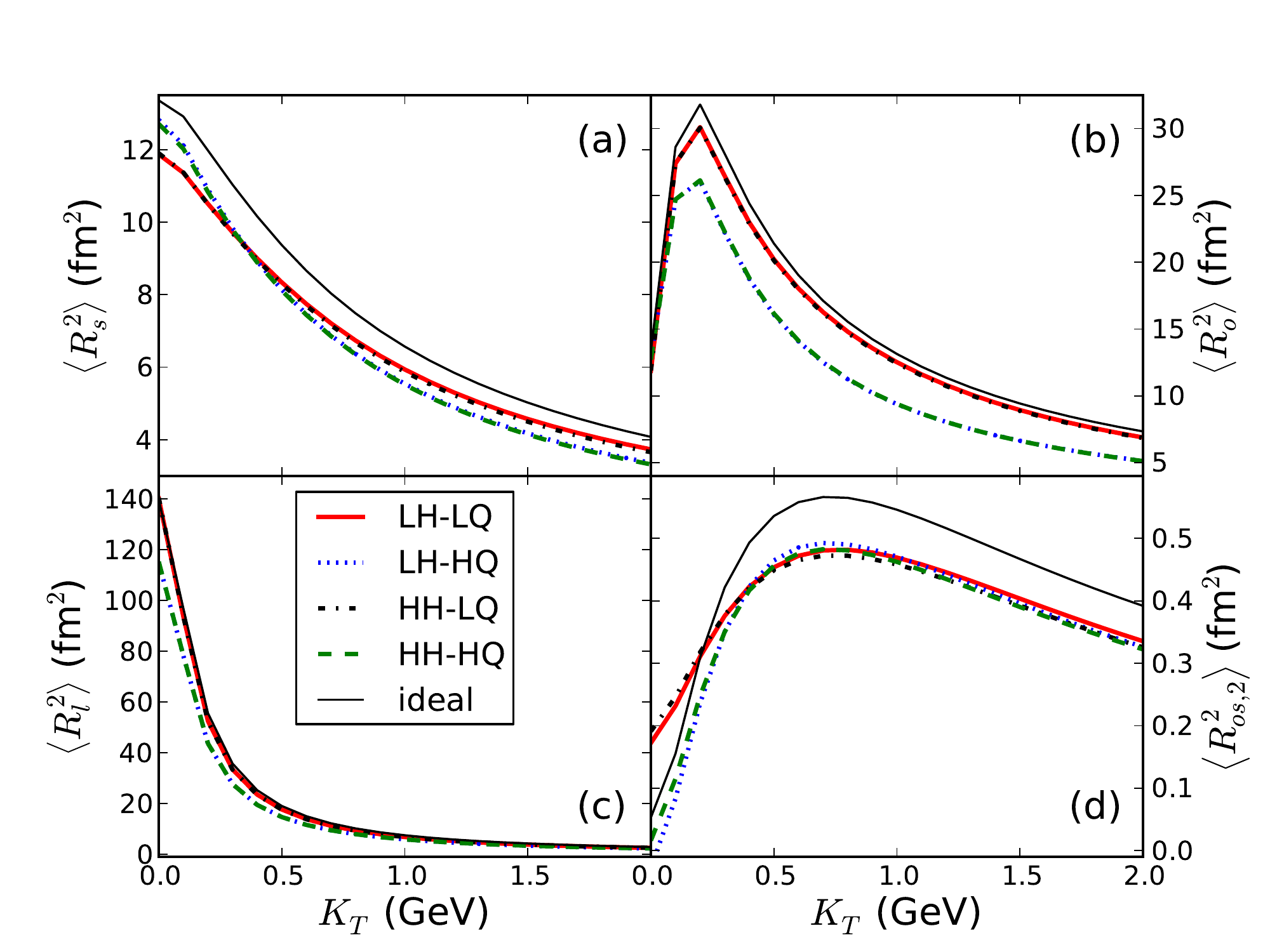}
\caption{Same as Fig.~\ref{F4}, but without the $\delta f$ correction. This figure should be compared with Fig.~\ref{F3}.
\label{F5}
}
\end{figure}

We now investigate how the mean HBT radii are affected by event-by-event fluctuations in the initial conditions. To do so we evolve 1000 central (0-10\% centrality) Au+Au events at $\sqrt{s}\eq200\,A$\,GeV with fluctuating initial conditions and average their correspondingly fluctuating HBT radii after freeze-out, as described in Sec.~\ref{ssec2d}. The results are presented in Fig.~\ref{F4} (including the full distribution function (\ref{cooper_frye_defn2}) at freeze-out) and Fig.~\ref{F5} (without the viscous $\delta f$ correction in (\ref{cooper_frye_defn2}) at freeze-out). Compared to Figs.~\ref{F2} and \ref{F3}, we see no qualitative differences. Again, robust (although not very strong) sensitivities to the temperature dependence of $\ebsT$ {\em above} $T_\mathrm{tr}$ are seen in all radii, but for $R_s^2$ these are buried by a stronger sensitivity to the viscous $\delta f$ correction at freeze-out (which, we reiterate, is theoretically not well controlled). As was the case for the $\bar{R}_{ij}^2$ from single-shot hydrodynamics, the $\delta f$ corrections to the $\mean{R_{ij}^2}$ are negligible when $i,j\eq{o,l}$.

%

\begin{figure}[t!]
\hspace*{-3.5mm}\includegraphics[width=1.05\linewidth]{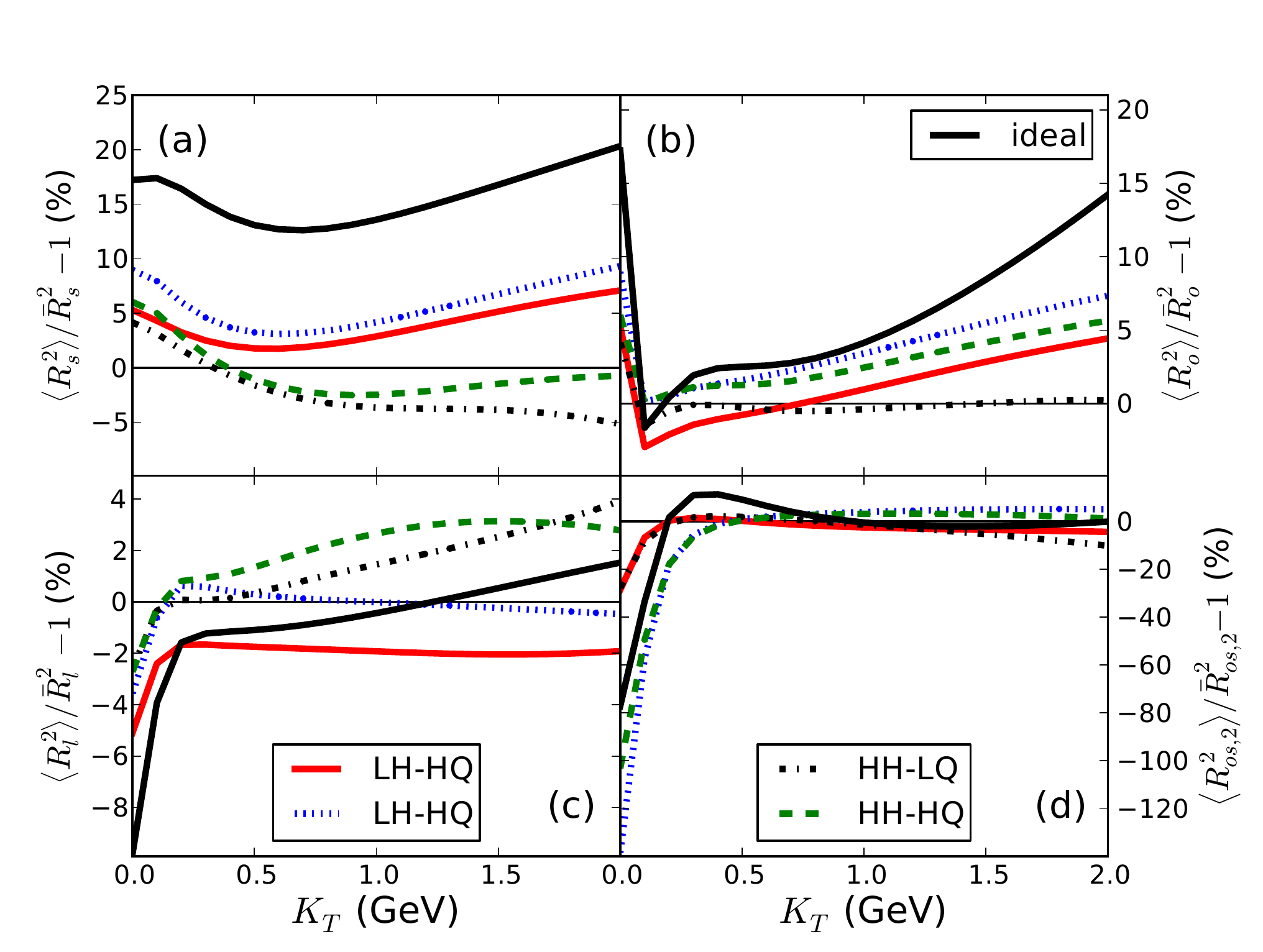}
\caption{The fractional change $\left(\mean{R_{ij}^2}{-}\bar{R}_{ij}^2\right)/\bar{R}_{ij}^2$ of the squared HBT radii due to event-by-event fluctuations, as a function of pair momentum $K_T$. See text for discussion.
\label{F6}
}
\end{figure}

In Fig.~\ref{F6} we plot the fractional change $\l(\mean{R_{ij}^2}{-}\bar{R}_{ij}^2\r)/\bar{R}_{ij}^2$ of the mean squared HBT radii arising from event-by-event fluctuations, as a function of pair momentum. Generically, event-by-event fluctuations are seen to boost the $\Phi_K$-averaged means of the fluctuating radii by a few percent above the corresponding radii obtained from single-shot hydrodynamic evolution of a smooth averaged initial condition. Exceptions to this rule are the sideward radius for the HH-LQ and HH-HQ parametrizations (i.e., for large shear viscosities at freeze-out) in the range $K_T{\,>\,}0.5$\,GeV, driven by effects from the $\delta f$ correction at freeze-out, and the longitudinal radius at small pair momentum $K_T{\,<\,}0.2$\,GeV. The systematic effects of event-by-event fluctuations on the mean $\Phi_K$-oscillation amplitude $\mean{R^2_{os,2}}$ are less unambiguous, with fluctuations increasing or decreasing $\mean{R^2_{os,2}}$ relative to $\bar{R}^2_{os,2}$ in different $K_T$-ranges depending on the specific temperature dependence selected for $\ebsT$; the differences can become substantial at small $K_T$ where the oscillation amplitude is small, with $\mean{R^2_{os,2}}$ even changing sign relative to $\bar{R}^2_{os,2}$ in the LH-HQ and HH-HQ parametrizations. Generically the differences between the mean fluctuating HBT radii and their single-shot hydrodynamic analogs are largest for ideal fluid evolution and somewhat smaller for viscous evolution. This is consistent with the idea that viscosity attenuates fluctuation-driven effects. 

\begin{figure*}
	\begin{tabular}{@{}c@{}}
		\includegraphics[scale=0.62]{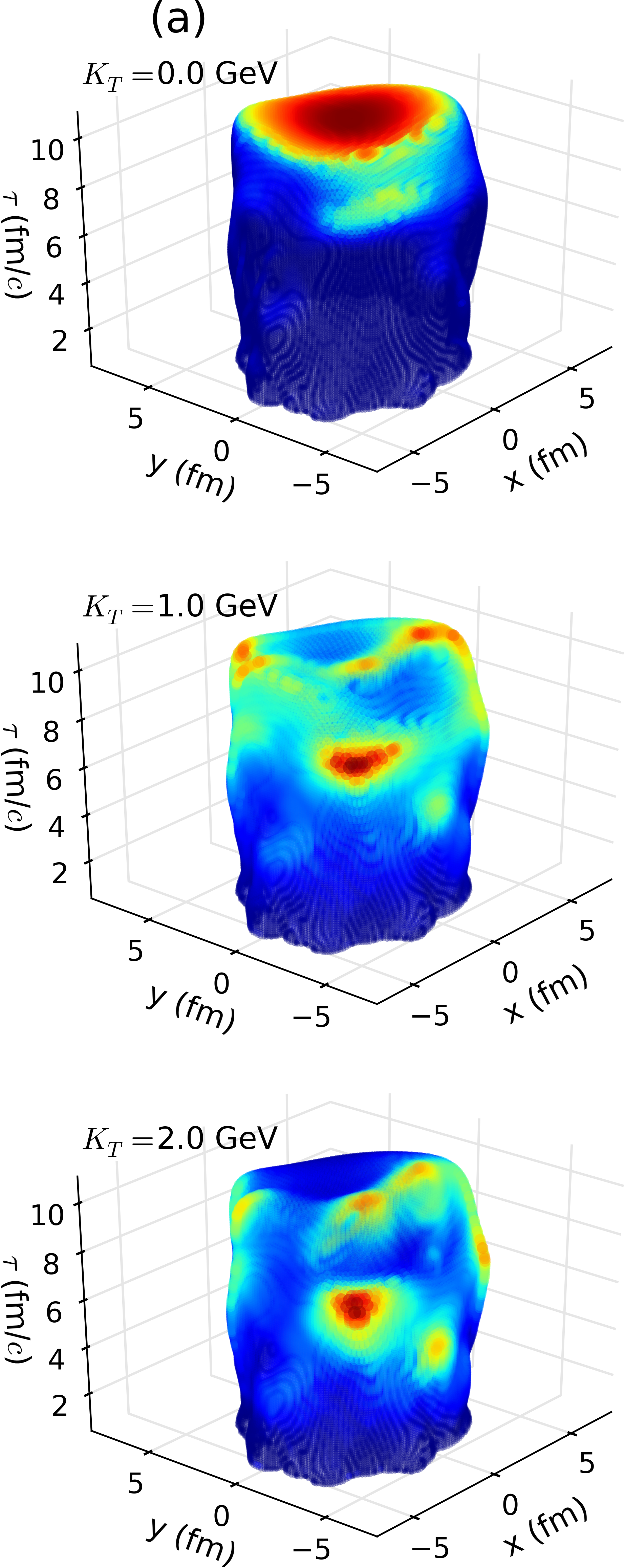}
		\qquad
		\includegraphics[scale=0.68]{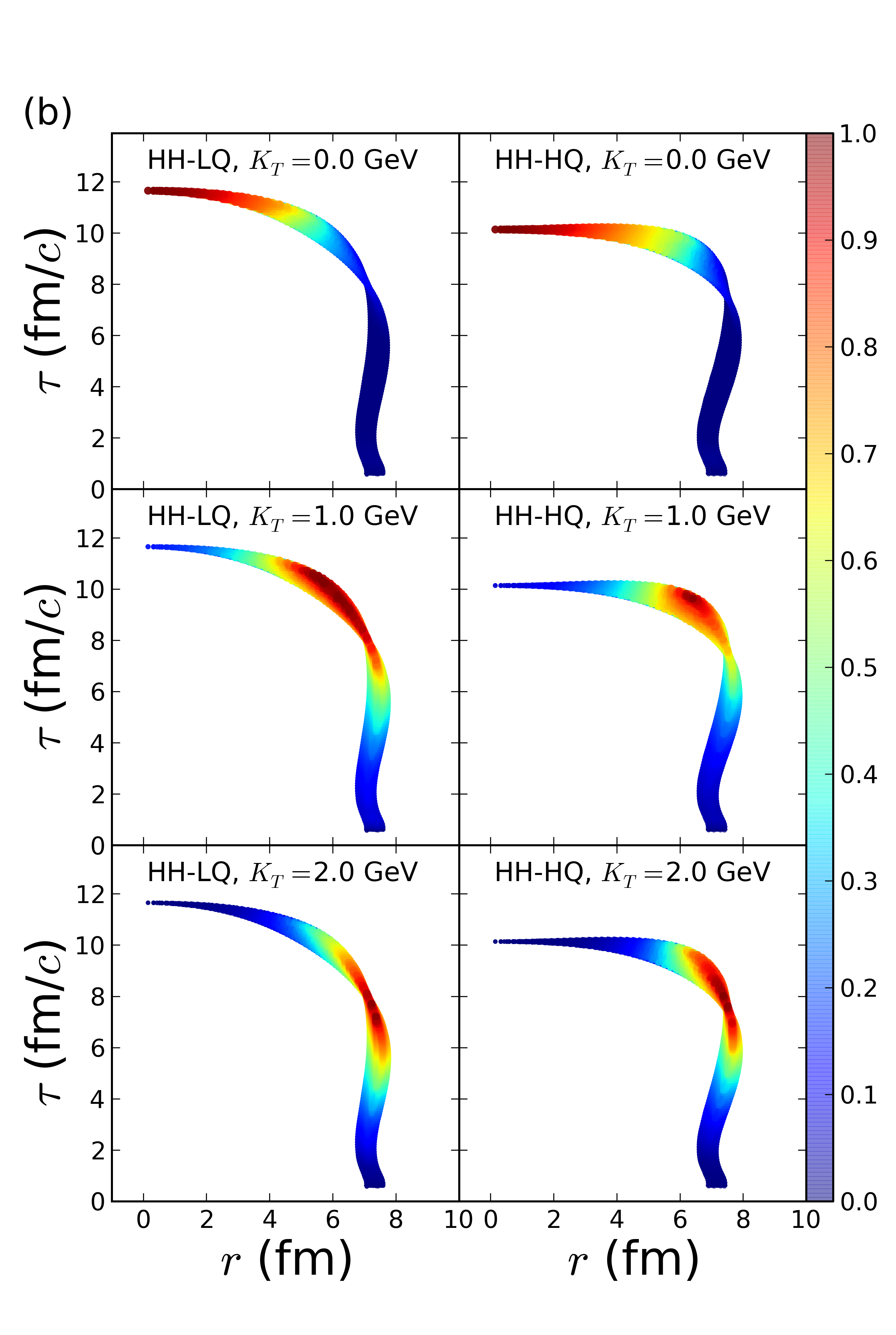}%
	\end{tabular}
\caption{(a) The emission function $S(x,K)$, for three values of $K_T$ as indicated, $Y\eq0$ and integrated over $\Phi_K$ and $\eta_s$, for a single fluctuating event from the 0-10\% centrality range, evolved with $(\eta/s)_\mathrm{HH-LQ}$.  Colors code the emission intensity, normalized to its maximum value on the freeze-out surface. (b) The emission function $S(x,K)$ for the smooth average single-shot hydrodynamic event corresponding to 0-10\% centrality, for the same three values of $K_T$, $Y\eq0$ and integrated over $\Phi_K$ and $\eta_s$, using the same color code as in (a). The width of the bands arises from plotting the ($\tau$, $r$) freeze-out contours for all spatial angles $\phi_s$ on top of each other. The left and right columns of panels show results for evolution with $(\eta/s)_\mathrm{HH-LQ}$ and $(\eta/s)_\mathrm{HH-HQ}$, respectively. See text for discussion.
\label{F7}
}
\end{figure*}

\begin{figure*}
\includegraphics[width=.7\textwidth]{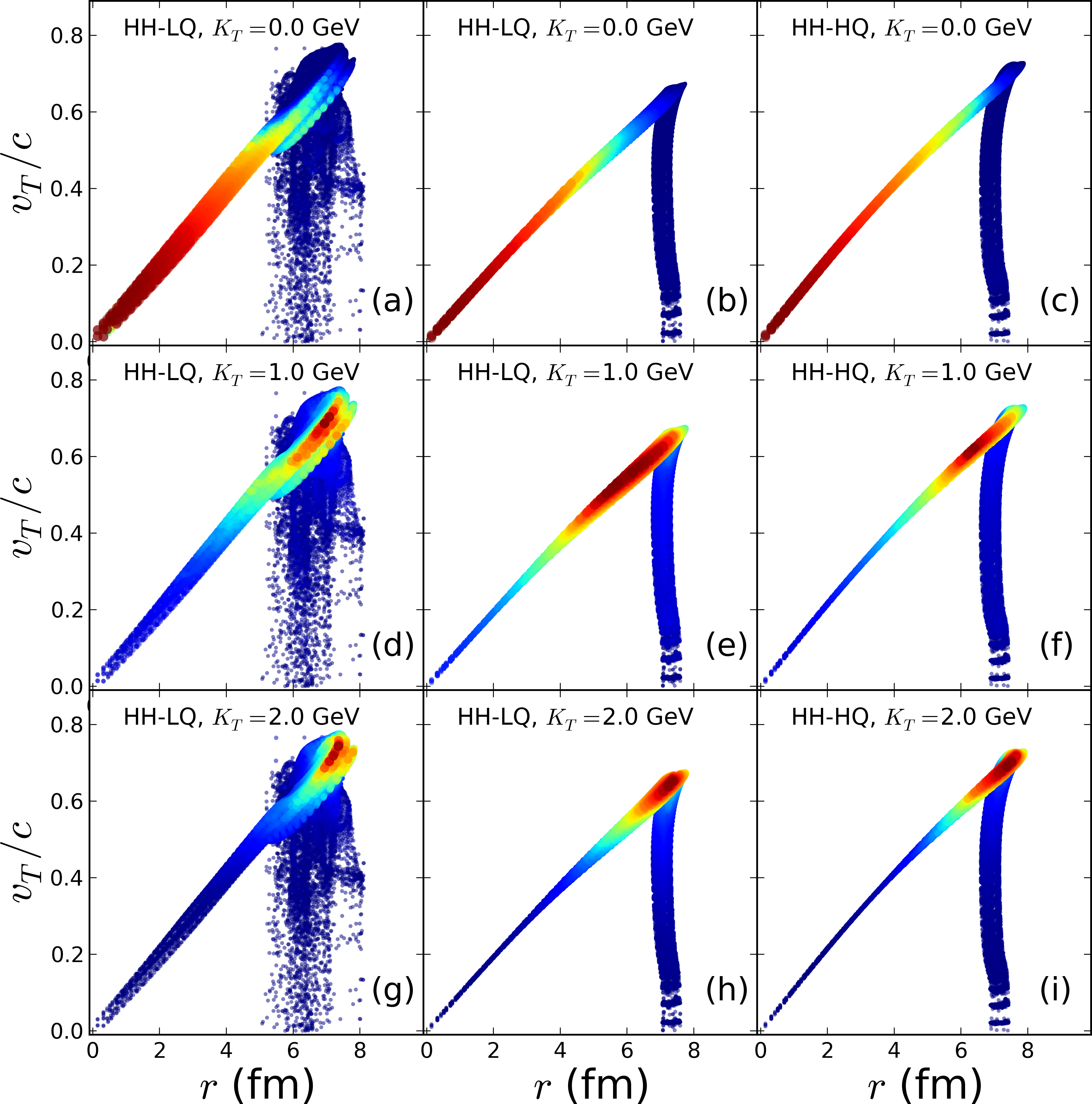}
\caption{For each cell on the freeze-out surface the radial velocity is plotted against its radial position, color coded as in Fig.~\ref{F7} for its emission intensity normalized by its maximal value on the surface. Panels a,d,g in the left column are for the single fluctuating event, evolved with $(\eta/s)_\mathrm{HH-LQ}$, shown in Fig.~\ref{F7}a, for the same three $K_T$ values as indicated. Panels b,e,h in the middle column and panels c,f,i in the right column are for the ensemble-averaged event shown in Fig.~\ref{F7}b, evolved with single-shot hydrodynamics using $(\eta/s)_\mathrm{HH-LQ}$ (middle column) and $(\eta/s)_\mathrm{HH-HQ}$ (right column), respectively. See text for discussion.
\label{F8}
}
\end{figure*}

For a better understanding of how different temperature dependences of $\ebsT$ affect the magnitudes and $K_T$ dependences of the radius parameters $\bar{R}^2_{ij}$ and $\mean{R^2_{ij}}$ we can look directly at the emission functions. These are shown in Fig.~\ref{F7}a for a randomly selected single fluctuating event, and in Fig.~\ref{F7}b for the single-shot hydrodynamic event with smooth ensemble-averaged initial conditions. Having seen in Figs.~\ref{F2}-\ref{F5} that at RHIC energies the HBT radii discriminate mostly between $\ebsT$ parametrizations that differ in the value of $\eta/s$ at the earliest times and highest temperatures, we focus in Fig.~\ref{F7}b on the HH-LQ and HH-HQ parametrizations (a comparison of LH-LQ with LH-HQ would lead to similar conclusions). 

As is well known (see, e.g., Ref.~\cite{Teaney:2001av}), the average radial flow velocity increases approximately linearly with $r$ as one moves from the top of the freeze-out surface in Fig.~\ref{F7}b (at $r\eq0$) to its vertical part (where $r\eq{r}_\mathrm{max}$ takes its maximum value). For the cases studied in Fig.~\ref{F7}, this is shown in Fig.~\ref{F8}. This observation explains why, for increasing $K_T$, the region of maximal emissivity (colored red) moves from around $r\eq0$ for small momentum pairs to $r_\mathrm{max}$ for large momentum pairs. Fig.~\ref{F7}a illustrates how this phenomenon manifests itself in an individual fluctuating event: While the regions of highest emissivity roughly follow the average tendencies seen in Fig.~\ref{F7}b, these tendencies are strongly modulated by fluctuations in the shape of the freeze-out surface and of the radial flow along this surface, giving rise to peaks and valleys of emissivity as one moves around the freeze-out surface at constant values of $r$ or $\tau$. Fig.~\ref{F7}a also illustrates that, in an individual event, the regions of maximal emissivity have a strong dependence on the azimuthal direction $\Phi_K$ of pair emission, and that therefore, for an individual fluctuating event, one should expect strong $\Phi_K$-dependence of the HBT radius parameters.

Comparing the freeze-out surfaces and regions of maximal emissivity shown in Fig.~\ref{F7}b for the two ensemble-averaged sources that were evolved with shear viscosities $\ebsT$ differing only in their behavior at high temperatures, we can make several observations: \\
(1) The larger specific shear viscosity of the HH-HQ parametrization drives stronger radial flow which causes the fireball to expand to larger radii and complete its freeze-out at earlier times than for the HH-LQ parametrization. This agrees with similar observations made in Ref.~\cite{Song:2007ux}. \\
(2) For longitudinally boost-invariant expansion with longitudinal velocity component $v_z\eq{z}/t$, emissivity regions centered at later times see a smaller longitudinal flow gradient and thus a larger longitudinal region of homogeneity, reflected in a larger value for $R_l^2$ \cite{Heinz:1999rw}. Fig.~\ref{F7}b shows that  maximal emissivity for small and intermediate $K_T$ pairs is shifted to later times for the HH-LQ parametrization than for HH-HQ, while for large $K_T\eq2$\,GeV the pairs are emitted at roughly the same times for both parametrization. This explains why in Figs.~\ref{F2}c (and also in Fig.~\ref{F4}c) the longitudinal HBT radii are larger for HH-LQ than for HH-HQ at small and intermediate $K_T$ but nearly identical at large $K_T$.\\
(3) For small $K_T$, the regions of highest emissivity are geometrically smaller in the outward and sideward%
\footnote{%
    To see the truth of this second statement the reader should in her mind visualize the emission
    surface rotated in $\phi$ direction and look at its horizontal width when frontally viewed.
    }
directions for the HH-LQ case than for HH-HQ, but the opposite is true at intermediate and high pair momentum. This explains the behavior of the sideward radius $R_s^2$ for these two parametrizations shown in Figs.~\ref{F2}a.
 
The most important aspect of the freeze-out surfaces shown in Figs.~\ref{F7}b is the earlier freeze-out associated with larger shear viscosities at early times (high temperatures). It reduces the emission duration and increases the longitudinal flow gradient at freeze-out, which reduces the $\avg{\tilde{x}_l^2}$ and $\avg{\tilde{t}^2}$ contributions to $R_l^2$ and $R_o^2$ in Eqs.~(\ref{R2sl_defn}), (\ref{R2o_defn}) and is the root cause for the robustly smaller $R_o^2$ and $R_l^2$ values observed in Figs.~\ref{F2}b-c and \ref{F4}b-c for the HH-HQ parametrization of $\ebsT$ compared to the HH-LQ case. Its effect on the cross-terms variances $\avg{\tilde{x}_o\tilde{x}_s}$, $\avg{\tilde{x}_o\tilde{t}}$ and $\avg{\tilde{x}_s\tilde{t}}$ (and thus the behavior of the cross-term radius $R_{os}^2$) is less intuitively obvious.

Last but not least, it is well established that the rate at which $R_s^2$ decreases with increasing pair momentum $K_T$ is a measure of the radial velocity gradient along the freeze-out surface \cite{Heinz:1999rw}. Comparison of Figs.~\ref{F8}b,e,h (middle column) to Figs.~\ref{F8}c,f,i (right column) shows that evolution with $(\eta/s)_\mathrm{HH-HQ}$ leads to a larger radial flow gradient along the freeze-out surface than evolution with $(\eta/s)_\mathrm{HH-LQ}$ (which has smaller shear viscosity at early times). This correlates with the steeper $K_T$ dependence for $R_s^2$ in Figs.~\ref{F2}a and \ref{F4}a for the HH-HQ parametrization (green dashed curves) than for the HH-LQ one (black dash-dotted curves).

\subsection{Variances of HBT radii and their $K_T$ dependence}
\label{sec3b}

\begin{figure}[t!]
\hspace*{-3.5mm}\includegraphics[width=1.05\linewidth]{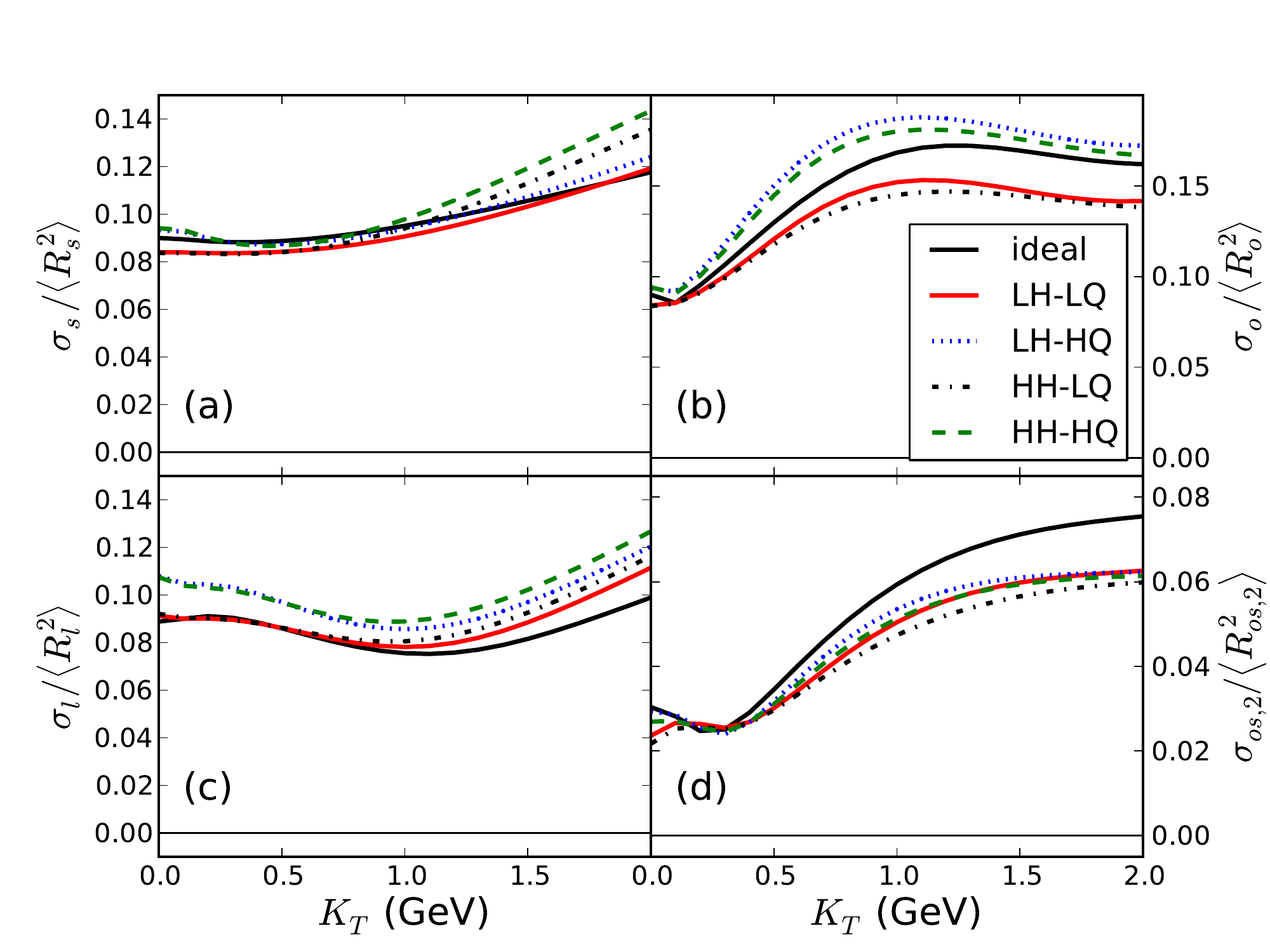}
\caption{ The normalized widths of the event-by-event distributions of HBT radii as functions of pair momentum $K_T$, for ideal fluid evolution and viscous dynamics with the four parametrizations shown in Fig.~\ref{F1}.
\label{F9}
}
\end{figure}

\begin{figure}[!t]
\hspace*{-3.5mm}\includegraphics[width=1.05\linewidth]{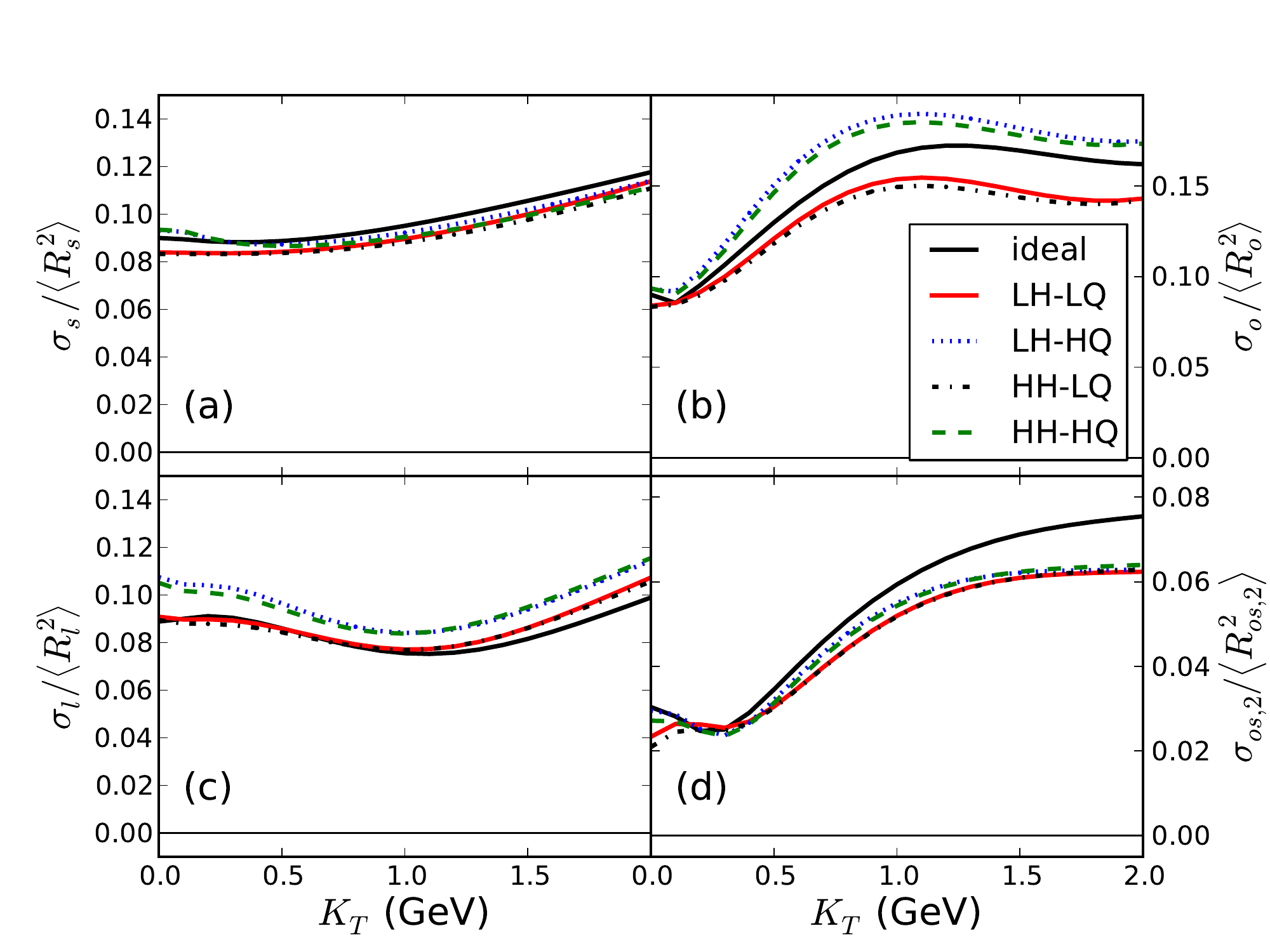}
\caption{Same as Fig.~\ref{F9}, but without the viscous correction $\delta f$ to the freeze-out distribution function. One sees that for the normalized width of the distribution of sideward radii $R_s^2$, deleting $\delta f$ removes all sensitivity to the temperature dependence of $\eta/s$ in the hadronic phase, while all other normalized widths are only weakly affected by $\delta f$.
\label{F10}
}
\end{figure}

Finally, we consider the variances of the event-by-event distributions of the HBT radii.  We present them in the form of \textit{relative} widths, normalized to the directly ensemble averaged squared radii, in Figures~\ref{F9} (full freeze-out distribution function) and \ref{F10} (without the viscous $\delta f$ correction on the freeze-out surface). For the $\Phi_K$-averaged radius parameters the relative widths are seen to be of order $10-15\%$, with little dependence on either $K_T$ or the specific shear viscosity used in the hydrodynamic evolution. The event-by-event distribution of the out-side oscillation amplitude $R_{os,2}^2$ shows a somewhat smaller relative width, especially at small $K_T$ where the mean amplitude is small. At larger $K_T$, shear viscosity appears to reduce the relative width of the event-by-event distribution of this  oscillation amplitude compared to ideal fluid evolution.

Focusing on the $\Phi_K$-averaged radii (panels (a-c) in Figs.~\ref{F9} and \ref{F10}) we observe that, just as was the case for their ensemble averages, all sensitivity of their variances to the temperature dependence of $\eta/s$ in the hadronic phase appears to come from the viscous $\delta f$ correction at freeze-out and thus not to be of dynamical origin. The sensitivity of the HBT variances to dynamical evolution effects caused by varying the temperature dependence of $\eta/s$ at temperatures above freeze-out can be seen in Fig.~\ref{F10} where the $\delta f$ contribution at freeze-out is removed. The results for viscous evolution split into two bands, one for high, the other for low shear viscosity at early times. The relative widths for ideal fluid dynamics fall in between these two bands. The two bands are characterized by the behavior of $\eta/s$ at {\em high} temperature, not at low $T$: Higher shear viscosity at early times (i.e. at high temperatures) leads to HBT variances that are 10-20\% larger than those for evolution with lower shear viscosity at early times. The behavior of the shear viscosity at late times in the hadronic phase has no visible dynamical effect on the normalized widths of the HBT radii.

\section{Conclusions}
\label{sec4}

We have presented a first analysis of the sensitivity of the HBT radii to the temperature dependence of the specific shear viscosity in relativistic heavy ion collisions at top RHIC energies, taking into account that the HBT radii fluctuate from event to event and that a consistent comparison of experimental data with theoretical predictions should thus be based on event-by-event evolution of fluctuating initial conditions on the theory side. We explored the effect of event-by-event fluctuations on the mean HBT radii (and their deviation from the values obtained from single-shot hydrodynamics where the initial-state fluctuations are ignored by averaging over them before hydrodynamic evolution) and their relative widths. We found that fluctuations tend to increase the mean value of the fluctuating HBT radii by a few percent above the value obtained by the traditional process of averaging over the fluctuations already in the initial state, and found that the event-by-event distributions of the HBT radii are characterized by relative widths of order 10\%. 

Comparing the HBT radii from hydrodynamic evolutions with five different assumptions for the specific shear viscosity and its temperature dependence we showed that both the mean squared HBT radii and their relative widths are affected by a possible increase of the specific shear viscosity by a factor 10 between $T_\mathrm{tr}$ and $3\,T_\mathrm{tr}$ at the level of 10-20\%, with larger shear viscosities leading to smaller mean HBT radii and larger relative widths of their event-by-event distributions. There is little to no effect on either from variations in the behavior of $\ebsT$ in the hadronic phase below $T_\mathrm{tr}$, except for the sideward and out-side radius parameters $R_s^2$ and $R^2_{os,2}$ and the sideward variance $\sigma^2_s$ which all show significant sensitivity to the viscous $\delta f$ correction at freeze-out. The latter, in turn, depends on the value of $\eta/s$ at the freeze-out temperature but not on its prior dynamical evolution in the hadronic phase. Since the $\delta f$ correction is not theoretically well constrained at the present moment, this sensitivity of $R^2_s$ and $R^2_{os,2}$ to $\eta/s$ at freeze-out should be viewed with some caution.  

The levels of sensitivity of the HBT radii and their variances to the temperature dependence of $\eta/s$ observed here are comparable to those observed in Ref.~\cite{Molnar:2014zha} for the elliptic and quadrangular flows. This indicates that precise 3-dimensional HBT measurements, in particular new measurements of their variances \cite{Plumberg:2015cjp}, can play a valuable supporting role in constraining the temperature dependence of the QGP shear viscosity. 

\vfill

\acknowledgments
The authors would like to thank Chun Shen for a careful reading of the manuscript which exposed an error in its original version. This work was supported by the U.S. Department  of Energy, Office of Science, Office of Nuclear
Physics under Awards No. \rm{DE-SC0004286} and (within the framework of the JET Collaboration) \rm{DE-SC0004104}.



\end{document}